\documentclass{llncs}

\usepackage{pslatex}
\usepackage{pifont}
\usepackage{epsfig}
\usepackage{graphics}
\usepackage{stmaryrd}
\usepackage{color}
\usepackage{wrapfig}
\usepackage{subfigure}
\usepackage{latexsym}
\usepackage{float}
\usepackage{listings}
\usepackage{graphicx}

\newcommand{\ignore}[1]{}

\newcommand{\false}{\mbox{\frenchspacing \it false}}

\newcommand{\clpif}{\mbox{\tt :-}}

\newcommand{\wrt}{{\frenchspacing wrt }}

\newcommand{\dido}[1]{}

\newcommand{\lbr}{\mbox{$[\![$}}
\newcommand{\rbr}{\mbox{$]\!]$}}

\newcommand{\pp}[1]{\mbox{{$\langle$#1$\rangle$}}}

\newcommand{\G}{\mbox{$\cal G$}}

\newcommand{\Gone}{\G_{\!\!1}}

\newcommand{\mymax}{\mbox{\frenchspacing \it max}}
\newcommand{\mymin}{\mbox{\frenchspacing \it min}}

\newcommand{\myquote}[1]{''\emph{#1}''}

\newcommand{\dpred}[1]{\mbox{\textit{#1}}}
\newcommand{\dspred}[1]{\mbox{\scriptsize\textit{#1}}}

\newcommand{\mysection}[1]{\vspace{-1mm} \section{#1} \vspace{-1mm}}

\begin{document}

\title{Symbolic Execution For Verification}

\author{\sc Joxan Jaffar, Jorge A. Navas, and Andrew E. Santosa}

\institute{ National University of Singapore\\
 {\tt \{joxan,navas,andrews\}@comp.nus.edu.sg} \\
}
\maketitle

\begin{abstract}

  In previous work, we presented a symbolic execution method which
  starts with a concrete model of the program but progressively
  abstracts away details only when these are known to be irrelevant
  using interpolation.  In this paper, we extend the technique to
  handle unbounded loops.  The central idea is to progressively
  discover the strongest invariants through a process of loop
  unrolling.  The key feature of this technique, called the minimax
  algorithm, is \emph{intelligent backtracking} which directs the
  search for the next invariant.  We then present an analysis of the
  main differences between our symbolic execution method and
  mainstream techniques mainly based on abstract refinement
  (\textsc{cegar}).  Finally, we evaluate our technique against
  available state-of-the-art systems.

\end{abstract}

\vspace*{-0.75cm}

\mysection{Introduction}
\label{sec:intro}

\emph{CounterExample-Guided Abstraction Refinement}
(\emph{\textsc{cegar}\/}, or more briefly,
\emph{\textsc{ar}\/})~\cite{Clarke00counterexample-guidedabstraction,ball01predicate,SaidiSAS00},
has been a very successful technique for proving safety in large
programs. Starting with a coarse abstraction of the program
(\emph{abstraction phase}), the abstraction is checked for the desired
property (\emph{verification phase}). If no error is found, then the
program is safe. Otherwise, an abstract counterexample is produced.
The counterexample is then analyzed to test if it corresponds to a
concrete counterexample in the original program. If yes, the program
is reported as unsafe. Otherwise, a \emph{counterexample-driven
  refinement} is performed to refine the abstract model such that the
abstract counterexample is excluded (\emph{refinement phase}), and the
process starts again. Several systems have been developed during
recent years following this
approach~\cite{Ball04slamand-short,chaki04magic,henz02lazy,ARMC07-short,cousot07fixpoint,SynergyFSE06,DASHISSTA08,SMASH-POPL10}.

In a previous work~\cite{jaffar09intp} we presented a \emph{dual\/}
algorithm to \textsc{ar}, here called \emph{Abstraction Learning}, for
loop-free program fragments. Essentially, our technique starts with
the concrete model of the program. Then, the model is checked for the
desired property (\emph{verification phase}) via \emph{symbolic
  execution}. If a counterexample is found, then it must be a real
error and hence, the program is unsafe. Otherwise, the program is
safe. In order to make the symbolic execution process practical, the
technique learns the facts that are irrelevant for keeping infeasible
paths by computing \emph{interpolants} (\emph{learning phase}), and
then it eliminates those facts from the model (\emph{abstraction
  phase}). Unfortunately, this work did not provide an automatic
treatment of loops while it assumed user-provided loop invariants to
make symbolic executions finite.

In this paper, we extend the technique proposed in~\cite{jaffar09intp}
to discover loop invariants. The central idea is to progressively
discover the strongest invariants through a lazy process of loop
unrolling.

For a given loop, \emph{path-based loop invariants} are computed and
used to generalize the states at the looping points (program points
where the merging of control paths construct some cyclical paths). Our
computation of invariants is {\em lightweight\/} as they are computed
by manipulation, using the theorem prover, of explicit
constraints. The algorithm attempts to minimize the loss of
information by computing the \emph{strongest} possible invariants.
These \emph{speculative} invariants may be still too coarse to ensure
safety. Here the algorithm computes interpolants to ensure that error
locations are not reachable, resulting in \emph{selective} unrolling
at points where the path-based invariant can no longer be produced due
to the strengthening introduced by the interpolants. Similar to
\textsc{ar}, this procedure is only guaranteed to terminate when loop
iterations are bounded.

A fundamental distinction with \textsc{ar} is that we attempt
to always construct the most precise abstraction for loops by
computing the strongest lightweight loop invariants. This feature
is vital to detect as many infeasible paths as possible during the
symbolic execution-based traversal. Our thesis is that this investment
often pays off, and even in examples where it does not, it is
affordable.

The contributions of this paper can be summarized as follows:

\begin{enumerate}

\item We extend the interpolation-based symbolic execution algorithm
in~\cite{jaffar09intp} to deal with unbounded loops by describing a
novel lazy loop unrolling algorithm called \emph{minimax}.
 
\item We provide an analysis using several academic examples of the
major differences between our proposed algorithm and mainstream
techniques mainly based on abstraction refinement.

\item Finally, we implement the main ideas of this paper in a system
called \textsc{tracer}, and we evaluate it using real programs against
\textsc{blast}, available state-of-the-art system.

\end{enumerate}

\noindent \textbf{Related Work.} Our work is clearly related to
abstraction refinement
(\textsc{cegar})~\cite{Clarke00counterexample-guidedabstraction,ball01predicate,SaidiSAS00,henz02lazy,henzinger04proof}. We
dedicate Sec.~\ref{sec:overview} to exemplify main differences through
some academic examples and Sec.~\ref{sec:results} to compare with
\textsc{blast} using real programs.

Recent algorithms such as
Synergy/\textsc{dash}/\textsc{smash}~\cite{SynergyFSE06,DASHISSTA08,SMASH-POPL10}
use test-generation features to enhance the process of
verification. The main advantage comes from the use of lightweight
symbolic execution provided by \textsc{dart}~\cite{GodefroidKS05} to
mitigate the expensive cost of the \emph{abstract post-image operator}
when predicate abstraction is used. An advantage of our approach is
that it does not suffer from this drawback since ours is symbolic
execution-based and does not use predicate abstraction. More
importantly, these tools rely on \textsc{cegar} to build the abstract
model of the program, and hence, major limitations observed in
Sec.~\ref{sec:overview} still hold. Moreover, there is no benefit of
using test cases using our method unless there is a real
counterexample in the program. On the contrary, we can construct
reasonable scenarios where Synergy and its descendants can have an
exponential slowdown \wrt to ours as shown in Sec.~\ref{sec:overview}.

Our closest related works are in~\cite{jaffar08memo,jaffar09intp}.
where interpolation was performed on a search tree of a CLP goal in
pursuit of a target property.  (The earlier paper ~\cite{jaffar08memo}
focussed on a finite domain for an optimization problem.)  But these
works did not consider loops.  The main conceptual advance of this
paper is to address loops, and in doing so, allows for the
consideration of real-life programs.  Furthermore, this paper provides
a detailed analysis of differences with the state-of-the-art
\textsc{cegar} method, and finally, we present a comprehensive
experimental evaluation with \textsc{blast}, the most advanced
\textsc{cegar} implementation available to us at this time.

Very recently, another interpolation-based symbolic execution method
has been proposed, independent from ours,
in~\cite{mcmillan10lazy-short}. This work can be considered in two
parts.  In the consideration of loop-free program fragments, this work
is in fact subsumed by the earlier works
~\cite{jaffar08memo,jaffar09intp}.  In the consideration of loops,
~\cite{mcmillan10lazy-short} presented a \emph{naive} strategy for
handling loops based on an iterative deepening process.  The central
idea is to compute interpolants for a fixed depth in the hope they
will converge to inductive assertions after an expensive fixpoint
computation.  We quote from ~\cite{mcmillan10lazy-short}: \myquote{the
  question of how to obtain convergence in practice for unbounded
  loops needs further study}. Therefore, the description of a concrete
algorithm from this idealistic one is far from being
trivial. Furthermore, experimental evaluation was provided only in
regard to testing, and not for the case of verification.  In contrast,
in this paper we present a \emph{directed} approach which essentially
amounts to an intelligent backtracking strategy which takes into
account the reason for failure at the current stage.

\mysection{The Basic Idea}
\label{sec:basic_idea}

Our basic algorithm performs \emph{symbolic execution} of the programs
while attempting to find an execution path that reaches the
\textbf{error()} function. If such path cannot be found, then it
concludes that the program is safe.

\begin{figure}[t]
\begin{tabular}{lcc}
\hspace*{-0.75cm}
\begin{minipage}{0.2\textwidth}
\lstset{language=C, basicstyle=\ttfamily, lineskip=-5pt}
\begin{lstlisting}[morekeywords=error]
0: x=0;
1: if(*)
2:  x=x+1;
3: if(y>=1)
4:  x=x+2;
5: if(y<1)
6:  x=x+4;
7: if(x>5)
8:  error();
9:
\end{lstlisting}
\end{minipage} & 

\begin{minipage}[h]{0.5\textwidth}
  \includegraphics[height=5cm]{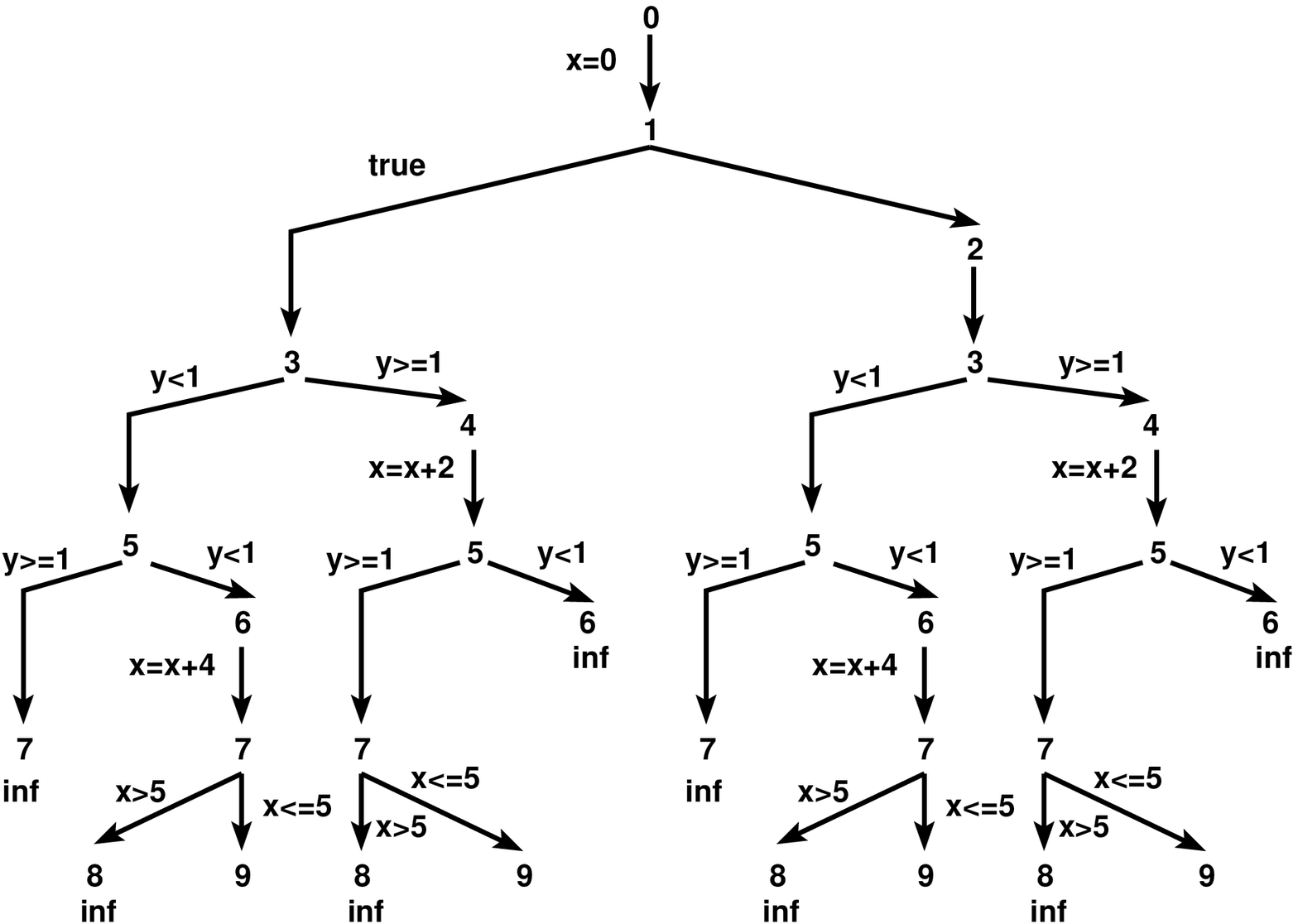}
\end{minipage}  &

\begin{minipage}[h]{0.5\textwidth}
\hspace*{1cm}
  \includegraphics[height=5cm]{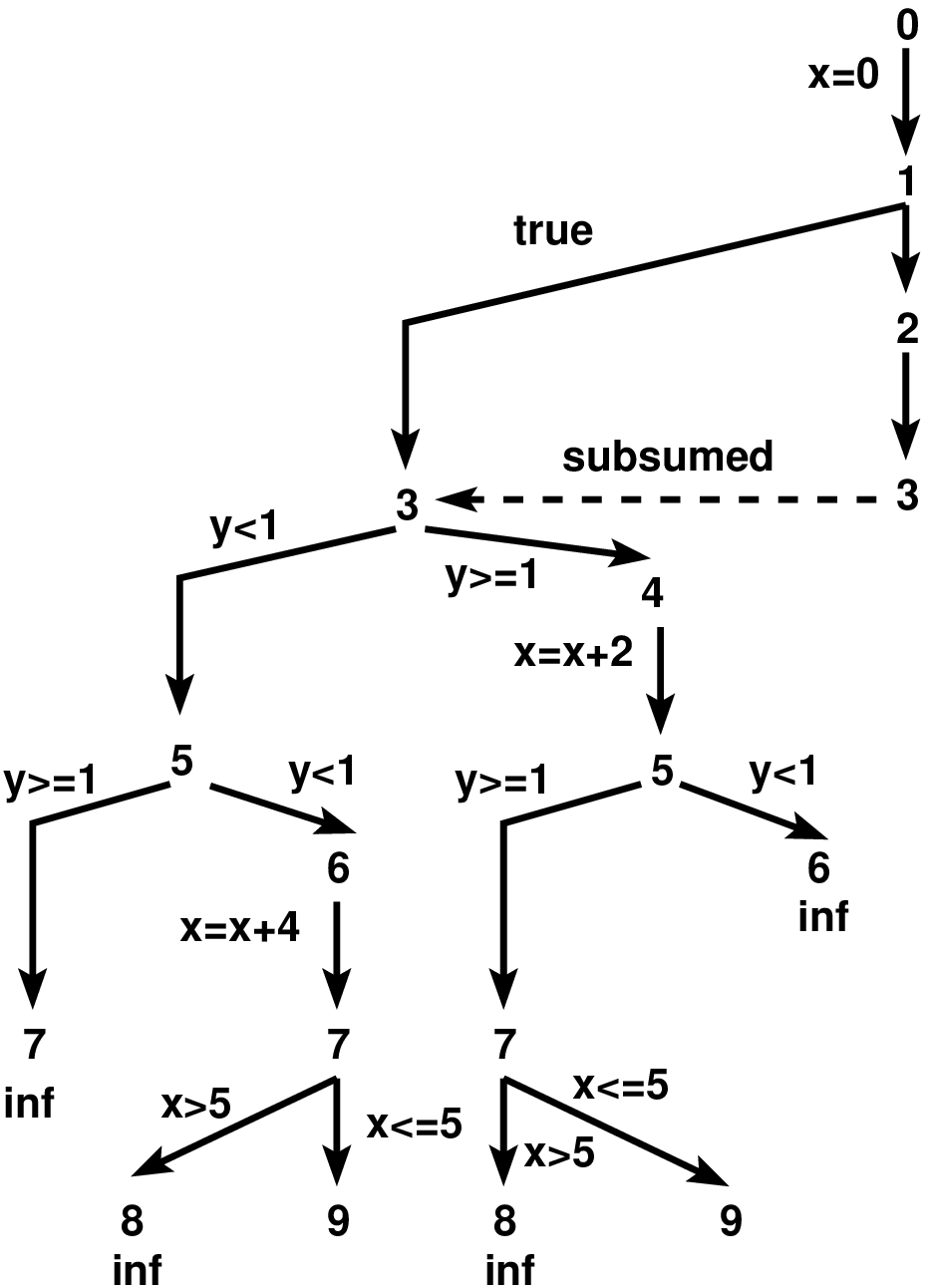}
\end{minipage} \\
\end{tabular}

\hspace*{0.3cm}  \mbox{(a)}  \hspace*{4cm} \mbox{(b)}  \hspace*{6cm} \mbox{(c)} \\
\vspace*{-0.5cm}
\caption{Interpolation and Subsumption of Infeasible Paths}
\label{fig:infeasible}
\vspace*{-0.5cm}
\end{figure}

Consider the program in Fig.~\ref{fig:infeasible}(a).  We depict in
Fig~\ref{fig:infeasible}(b) the naive symbolic execution tree, and in
Fig.~\ref{fig:infeasible}(c) a smaller tree, which still proves the
absence of bugs.  During the traversal of the tree, our algorithm {\em
  preserves the infeasibility\/} of the paths using the well-known
concept of \emph{interpolation}.  Let us focus on
Fig.~\ref{fig:infeasible}(c) and consider, for instance, the path $A
\equiv$ \pp{0}-\pp{1}-\pp{3}-\pp{5}-\pp{7} which is detected as
infeasible ($x=0 \wedge y < 1 \wedge y \geq 1$). Applying our
infeasibility preservation principle, we keep node \pp{7} labeled with
\false.  This produces the interpolant $y<1$ at node \pp{5} since this
is the most general condition that preserves the infeasibility of node
\pp{7}.  Note that here, $y<1$ is entailed by the original state $x=0
\wedge y<1$ of node \pp{5} and in turn entails $y \geq 1 \models
\false.$

Now consider another path $B \equiv$
\pp{0}-\pp{1}-\pp{3}-\pp{5}-\pp{6}-\pp{7}-\pp{8} and the node \pp{8}
with the formula $y<1 \wedge x=4 \wedge x > 5$ which is also
infeasible.  The node \pp{7} can be interpolated to $x \leq 5$. As
before, this would produce the precondition $x \leq 1$ at \pp{5}.  The
final interpolant for \pp{5} is the {\em conjunction} of $y<1$
(produced from $A$) and $x \leq 1$ (produced from $B$).  In this way,
when \pp{5} is visited through the path
\pp{0}-\pp{1}-\pp{3}-\pp{4}-\pp{5} the state cannot yet be subsumed
since the current context $y \geq 1 \wedge x=2$ does not entail the
interpolant stored at \pp{5} ($y<1 \wedge x \leq 1$). After that, the
symbolic execution continues normally until the the prefix $C \equiv$
\pp{0}-\pp{1}-\pp{2}-\pp{3} is traversed. The formula $x=0$ associated
to the state at \pp{3} entails the interpolant at $x\leq1$ at \pp{3}
and hence, our algorithm finishes proving safety without traversing
the whole subtree rooted at prefix $C$.

\begin{figure*}[t]
\vspace*{-1cm}
\begin{tabular}{lcc}

\begin{minipage}{0.3\textwidth}
\lstset{language=C, basicstyle=\ttfamily,lineskip=-5pt}
\begin{lstlisting}[morekeywords=error]
0:lock=0;
  new=old+1;
  flag=1;
1:while(new!=old){
2: lock=1;
   old=new;
3: if(*)
4:  lock=0;
    new++;
  }
5:if(!flag)
6: lock=0;
7:if(lock==0)
8: error();
9:  
\end{lstlisting}
\end{minipage}   &

\begin{minipage}[h]{0.5\textwidth}
  \includegraphics[height=4cm]{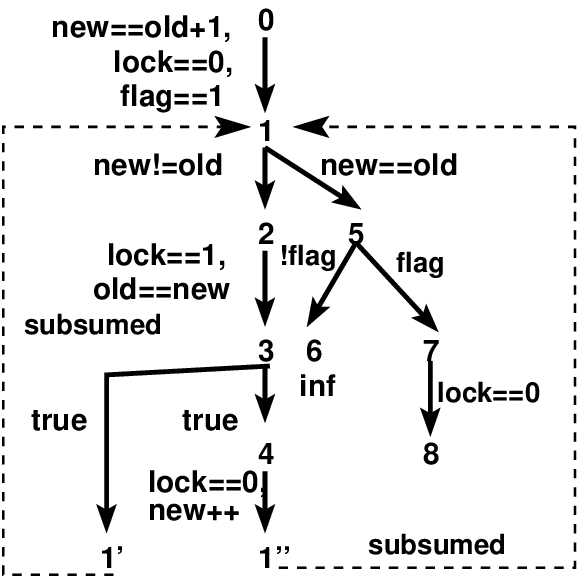}
\end{minipage} & 

\hspace*{-1.75cm}
\begin{minipage}[h]{0.5\textwidth}
\includegraphics[height=5.5cm]{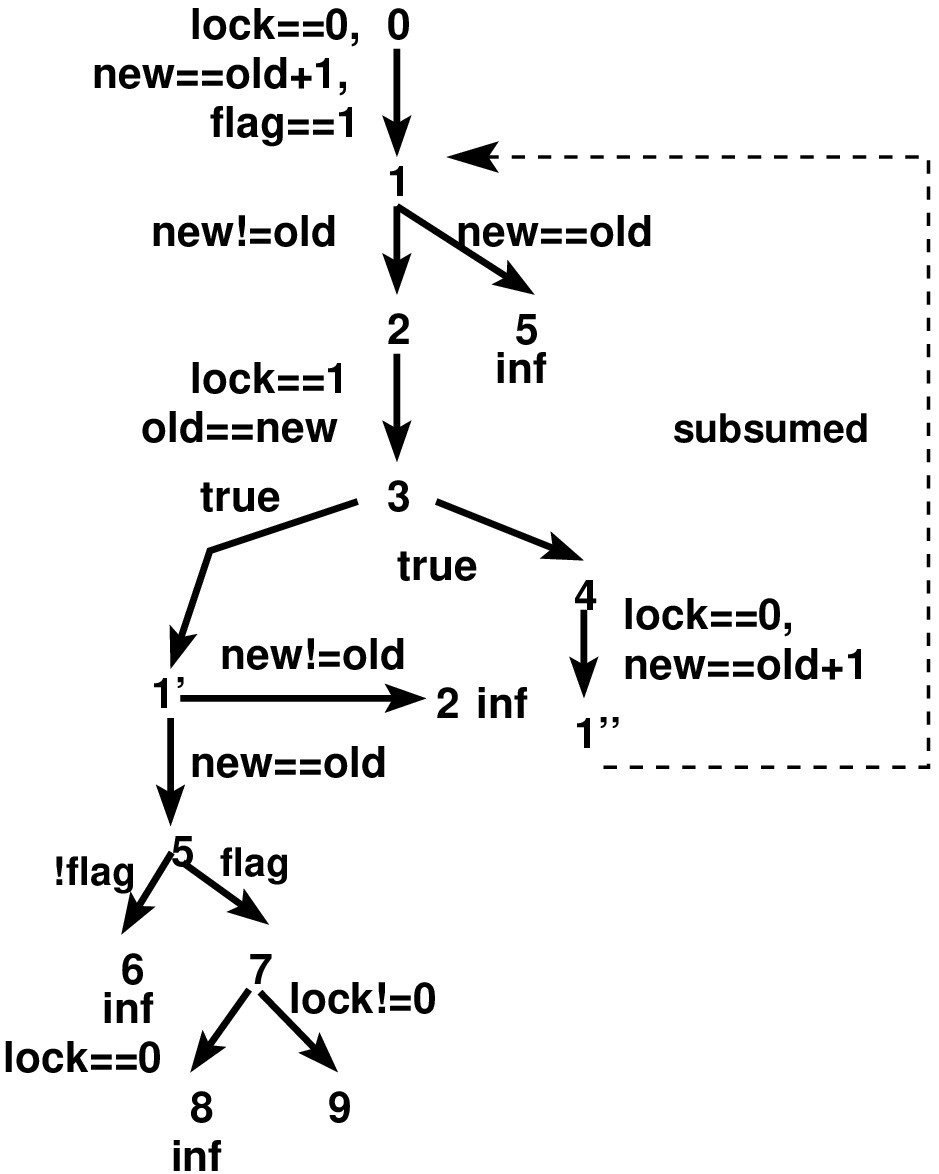}
\end{minipage} \\ \\
\hspace*{1cm} (a) & \hspace*{-0.5 cm} (b) & \hspace*{-2cm} (c) \\
\end{tabular}
\caption{Loops}
\label{example:loop}
\vspace*{-0.7cm}
\end{figure*}

\vspace*{0.25cm}
\noindent \textbf{Loops.} We now explain how our algorithm handles
loops using a slightly modified classic example from~\cite{henz02lazy}
shown in Fig.~\ref{example:loop}(a). Essentially, it automatically
infers path-based loop invariants using information learned during
traversal. The constructed loop invariant for a given path inside a
loop is a conjunction of constraints whose truth values remain
unchanged after one or more iterations of the loop. Similar to
abstraction refinement, this process may require refinements in the
case the abstraction is too coarse to prove the safety property.

In Fig.~\ref{example:loop}(b) assume the first path explored is
\pp{0}-\pp{1}-\pp{2}-\pp{3}-\pp{1'} denoting a cyclic path from
location \pp{1'} back to \pp{1}.  Note that \pp{1'} and \pp{1}
correspond to the same program point. We use primed versions to
distinguish multiple occurrences. Our algorithm then examines the
constraints at the entry of the loop (i.e., \texttt{lock==0},
\texttt{new==old+1},\texttt{flag==1}) to discover those whose truth
values remain unchanged after the loop (i.e.,
\texttt{lock==1}, \texttt{new==old}, \texttt{flag==1}). Clearly, the
constraints \texttt{lock==0} and \texttt{new==old+1} are no longer
satisfied while \texttt{flag==1} still holds.

At this point, our algorithm produces an abstraction at the location
\pp{1} by making the truth values of \texttt{lock==0} and
\texttt{new==old+1} unknown.  In this way, the constraints at \pp{1'}
now entails the modified constraints of \pp{1} (\texttt{flag==1}),
achieving parent-child subsumption.  Assume the next explored path is
\pp{0}-\pp{1}-\pp{2}-\pp{3}-\pp{4}-\pp{1''}
(Fig.~\ref{example:loop}(b)).  At \pp{1''}, the constraints already
entail the generalized constraint of \pp{1} (they are invariant), and
we therefore stop the traversal.

After the loop is traversed, the remaining constraint at \pp{1}
is \texttt{flag==1} and this is in fact a loop invariant discovered by
the algorithm.  Since we have removed \texttt{new==old+1} from
\pp{1}, the exit path of the loop now becomes feasible as the
condition \texttt{new==old} becomes satisfiable. For this reason the
traversal reaches \pp{5} with the constraint \texttt{flag==1}
propagated from \pp{1} and \texttt{new==old} which is obtained
by strongest postcondition propagation through the loop exit
transition.

Since we keep \texttt{flag==1} in the loop invariant at \pp{1}, the
algorithm manages to reason that the path \pp{0}-\pp{1}-\pp{5}-\pp{6}
is infeasible (Fig.~\ref{example:loop}(b)).  One important point here
is that the algorithm exits the loop with maximal information. This is
useful to detect as many infeasible paths as possible. An \textsc{ar}
algorithm would not detect the infeasibility and would visit
\textbf{error}() at \pp{8}.

Next, our algorithm visits the nodes \pp{7} and \pp{8} also in
Fig.~\ref{example:loop}(b), which is an error location.  The path is
spurious, and the algorithm discovers using interpolation that one of
the reason for the reachability of this point is the removal of
\texttt{new==old+1} at \pp{1}.  The algorithm decides to {\em lock\/}
\texttt{new==old+1} at \pp{1} and restarts the traversal from
\pp{1}. Locking declares that the constraint cannot be removed for
generating loop invariant. This is our main mechanism to ensure
\emph{progress}.

The next traversal after the locking is depicted in
Fig.~\ref{example:loop}(c).  Similar to the first traversal, the path
\pp{1}-\pp{2}-\pp{3}-\pp{1'} is again re-traversed. At \pp{1'}, the
constraints do not entail the constraints of \pp{1} anymore.  Due to
locking of \texttt{new==old+1}, we are prevented from generating a
loop invariant, and hence, subsumption does not hold. As the result,
the traversal continues, and it is completed without visiting the
error program point at \pp{8}.

An essential observation is that due to its directed search for loop
invariants, the algorithm does not unroll the location at \pp{1''}
(Fig.~\ref{example:loop}(c)) since the state is already subsumed by
\pp{1} without the need to force any abstraction. A naive iterative
deepening algorithm (e.g., ~\cite{mcmillan10lazy-short}) would also
unroll that path, and hence, we can construct reasonable scenarios in
which this leads to an exponential explosion.

\mysection{Comparison with the State-Of-The-Art } 
\label{sec:overview}

We now analyze essential differences between our approach and
mainstream techniques which are mainly based on abstraction refinement
(\textsc{cegar}).

\vspace*{0.15cm} \noindent \textbf{Exploration of Infeasible Paths.}
The core idea of abstraction refinement is to use the most general
abstraction first, and refine later.  This causes the exploration of
\emph{infeasible paths} which \emph{stresses significantly} well-known
problems in \textsc{ar}.  First, the more predicates are considered in
the abstract model the more costly will be the verification
phase. Moreover, if predicate abstraction is used (e.g., \textsc{slam}
and \textsc{blast}) expensive abstract post-image and quantifier
elimination are needed. Finally, the cost of the refinement process
may be also prohibitive.

Because of the huge impact of exploring infeasible paths significant
research has been done recently. A partial solution has been the use
of \textsc{dart} in order to provide a symbolic execution engine in
Synergy-like
tools~\cite{SynergyFSE06,DASHISSTA08,SMASH-POPL10}. However, the
construction of the abstract model is still needed and the above
problems persist.
Furthermore, these tools may perform unnecessary refinements that may
create reasonable scenarios which lead to an exponential behavior.
Consider the program in Fig.~\ref{fig:example}(a). Assume that a
Synergy-like tool produces the test case
\pp{1}-\pp{2}-\pp{3}-\pp{7}-\pp{10}, and that the abstract model, with
no predicates, reaches the error through the path
\pp{1}-\pp{2}-\pp{3}-\pp{5}-\pp{6}-\pp{8}-\pp{9}. Then, it tries now
to produce new test cases by negating the first constraint which is
not in the common prefix (i.e., \texttt{x>0}) but it is unsatisfiable
since $x=0$. Therefore, it will likely add the predicate $x \leq 0$
which is irrelevant for proving safety. Our technique will traverse
the path through \pp{1}-\pp{2}-\pp{3}-\pp{7}-\pp{8} and produce the
interpolant $q=1$. The rest of paths will entail that interpolant, and
hence, the behavior will be linear on the size of the program.

\vspace*{0.15cm} \noindent \textbf{Discovering Loop Invariants.}
Any symbolic traversal method will have to eventually discover loop
invariants that are strong enough for the proof process to conclude
successfully.  In the case of \textsc{ar}, the abstract model is
refined from spurious counterexamples by discovering which predicates
can refute the error path, and in this process, they are \emph{hoped}
to be in fact invariant through loops.

A crucial observation is that the inference of invariant predicates
can speedup significantly the convergence of loops~\cite{BeyerHMR07}.
We therefore employ invariant discovery by searching for the
\emph{strongest} invariants. This principle is also in accordance with
our philosophy to perform concrete symbolic execution in order to
maintain exact information for loop-free fragments.

Fig.~\ref{fig:example}(b) illustrates the benefits of computing
strongest loop invariants.
\textsc{ar} will discover the predicates $(n=0), (n=1), \ldots,
(n=N-1)$ and also $(y=0), \ldots, (y=N)$, and hence full unrolling of
the loop is needed.
To understand why our approach avoids the full unrolling, the concept
of inference of path-based loop invariant constraints is
essential. Consider the path
\pp{1}-\pp{2}-\pp{3}-\pp{4}-\pp{5}-\pp{3'}. The state at \pp{3'} can
be specified by the constraint $y = 0 \wedge n=0 \wedge n<N \wedge
y'=y+1 \wedge n'=n+1$ on the variables $x'$ and $y'$. Our algorithm
will attempt to infer which constraints are individually invariant in
order to get parent-child subsumption (i.e., ``close'' the loop). It
is straightforward to see that $y \geq 0$ (by slackening $y=0$ to $y
\geq 0$) is invariant through the loop because when \pp{3} is first
visited, $y=0 \Rightarrow y\geq 0$ holds and after one iteration the
constraints $y \geq 0 \wedge y'=y+1$ still imply $y'\geq 0$.
The second essential step is when the exit condition is taken (i.e.,
\pp{1}-\pp{3}-\pp{6}) our technique will attach $n \geq N$ to all
invariant constraints (in this case $y \geq 0$) by computing strongest
postcondition. More importantly, those two constraints ($y\geq 0
\wedge n \geq N $) suffice to prove that the error condition $y+n < N$
is false. Therefore, we are done with only one iteration through the
loop.

\begin{figure}[t]
\begin{small}
\begin{tabular}{llll}

\begin{minipage}[t]{0.3\textwidth}
\begin{center}
\lstset{language=C, basicstyle=\ttfamily\scriptsize,lineskip=-5pt}
\begin{lstlisting}[morekeywords=error]
 1:q=1; 
 2:if(*) 
 3:  x=0;
   else  
 4:  x=1;
 5:if(x>0) 
 6:  y=0;  
   else  
 7:  y=1;
 8:if(q==0) 
 9:  error();
10:
\end{lstlisting}
\end{center}
\end{minipage}

&

\hspace*{-1.6cm}
\begin{minipage}[t]{0.3\textwidth}
\begin{center}
\lstset{language=C, basicstyle=\ttfamily\scriptsize,lineskip=-5pt}
\begin{lstlisting}[morekeywords=error]
1:assume(y == 0);
2:n = 0;
3:while (n < N){
4:  y++;
5:  n++;
  }
6:if (y+n < N)  
7:  error();
\end{lstlisting}
\end{center}
\end{minipage}

&

\hspace*{-1cm}
\begin{minipage}[t]{0.4\textwidth}
\begin{center}
\lstset{language=C, basicstyle=\ttfamily\scriptsize,lineskip=-5pt}
\begin{lstlisting}[morekeywords=error]
 1:s=0;
 2:if(*) z=0;
 3:else  z=999;
   // 1
 4:if(*) s++ 
 5:else  s+=2;
   ...
   // N
 6:if(*) s++ 
 7:else  s+=2;  
 8:if(s+z>2*N && z==0) 
 9:  error();
\end{lstlisting}
\end{center}
\end{minipage}

& 
\hspace*{-1.6cm}
\begin{minipage}[t]{0.3\textwidth}
\begin{center}
\lstset{language=C, basicstyle=\ttfamily\scriptsize,lineskip=-5pt}
\begin{lstlisting}[morekeywords=error]
 1:if(*){
 2: x=0; 
 3: y=0;}
 4:else{ 
 5: x=complex_func(); 
 6: y=0;}    
 7:s=x;
 8:t=y;  
 /*1*/  
 9:if(*){s++;t++;}
   ...
 /*N*/
10:if(*){s++;t++;}
11:if(t>N && s>N) error();     
\end{lstlisting}
\end{center}
\end{minipage}

\\
\hspace{1cm} \mbox{(a)} & \hspace{-0.5cm} \mbox{(b)} & \mbox{(c)} &   \mbox{(d)} 
\end{tabular}
\end{small}
\caption{Several Programs}
\label{fig:example}
\vspace*{-0.75cm}
\end{figure}

\vspace*{0.15cm} \noindent \textbf{Using Newly Discovered Predicates in Future Traversal.}
Another fundamental question in \textsc{ar}: after the set of
predicates required to exclude the spurious counterexample has been
discovered, how should those predicates be used in other paths?
Consider our next program in Fig~\ref{fig:example}(c).

A counterexample-guided tool will discover the predicates $(s=0),
(s=1), \ldots, (s=2*N)$. Then, it will either add $(z=0)$ or
$(z=999)$.  Assume that it first adds the predicate $(z=0)$. The key
observation is that all the paths that include $(z=999)$ (location
\pp{3}) will be traversed considering all the predicates discovered
from paths that included $(z=0)$ (location \pp{2}), and hence, the
traversal will be exponential.  

Our algorithm will basically perform the same amount of work for the
case that $z=0$ is considered. However, it traverses the paths that
include $z=999$ without consideration of the facts learnt from paths
that include $z=0$ since it only keeps track of the concrete state
collected so far (i.e., $s=0 \wedge z=999$). Then, after the path
\pp{1}-\pp{3}-\pp{4}-\ldots-\pp{6}-\pp{8} is traversed we can discover
in a straightforward manner that $z=999$ suffices to refute the error
state and hence, the rest of the paths will be subsumed. Notice that
\textsc{ar} will also discover the predicate $(z=999)$ after the
counterexample is found. The essential difference, for this class of
programs, is that the predicates discovered previously ($(s=0), (s=1),
\ldots, (s=2*N)$ ) are used, and hence, the traversal will be
significantly affected by them.

\vspace*{0.15cm} \noindent \textbf{Running an Abstract State Hampers Subsumption.}
The next example illustrates another potential weakness of \textsc{ar}
that is not present in our approach. Even if locality is well
exploited, the likelihood of subsuming the \emph{currently traversed}
state may be diminished because the state, being abstract, is too
coarse.
Consider now the program in Fig.~\ref{fig:example}(d). Assume
\texttt{complex\_func} returns always $0$.

In principle, a counterexample-guided tool will behave very similarly
as in the program in Fig.~\ref{fig:example}(c). Assume that the prefix
path \pp{1}-\pp{2}-\pp{3} is taken. It will then discover the
predicates $(x=0), (s=0), (s=1), \ldots, (s=N), (y=0), (t=0), (t=1),
\ldots, (t=N)$. Again, those predicates are likely to be used during
the exploration of the else-branch (\pp{4}-\pp{5}-\pp{6}).  However,
an essential difference with respect to program in
Fig.~\ref{fig:example}(c) is that although the discovered predicate
$(x=0)$ is taken into consideration, the abstract state cannot be
covered since it is too coarse assuming it does not consider lazily
the value returned by \texttt{complex\_func}, and hence, it does not
entail the predicate $(x=0)$.
In contrast, since our method does perform a systematic propagation of
the program state the value returned by \texttt{complex\_func} will be
captured and we will be able to entail the interpolant $x=0$. The main
consequence is that the state now will be subsumed.

\vspace*{0.15cm} \noindent \textbf{Unnecessary Detection of Infeasible Paths.}
So far we have illustrated scenarios where our approach behaves better
than \textsc{ar}. The advantage exploited in the preceding examples is
the preservation of infeasible paths while abstracting loops using the
strongest lightweight loop invariants.
Unfortunately, this characteristic might be an important downside if
the program can be proved safe even traversing infeasible paths since
all the work of generating interpolants for preserving infeasible
paths would be wasteful.

We claim that eager detection of infeasible paths even if they are not
relevant to the safety property is not limiting in practice.  The
reason is that many of the infeasible paths in real programs must be
considered anyway to block the error paths, and hence,
counterexample-guided approaches will also consider them although
lazily paying a higher price later on.
The results obtained by our prototype with real programs shown in
Sec.~\ref{sec:results} support strongly our view.

To elaborate even more this point let us consider a real program
\texttt{statemate}~\cite{malardalenbenchmark} used commonly for
testing \textsc{wcet} tools. The program is generated automatically
and its main feature is the huge amount of infeasible paths.
We try to build the worst possible scenario by instrumenting the
program and adding \texttt{x=0} at the first statement of the program
where \texttt{x} is a fresh variable, and then adding the condition
\texttt{if (x>0)} \textbf{error}() at the end.  An \textsc{ar} tool
should add only the predicate $(x=0)$ to prove that the program is bug
free. However, an actual evaluation using \textsc{blast} shows some
significant performance degradation as it may not always choose the
right predicate, resulting in $21$ predicates discovered in 74 seconds
on Intel 2.33Ghz 3.2 GB (our algorithm takes 88 seconds).  This
experiment exhibits the worst possible scenario for our approach and
also illustrates another potential limitation of \textsc{ar}. If the
abstract error path has more than one infeasibility reason, then
existing refinement techniques have difficulties in choosing the right
refinement. Synergy-like tools mitigate this problem but introduce
other challenges as discussed above.

\mysection{Formalities}
\label{sec:preliminaries}

Here we briefly model a program as a transition system and formalize
the proof process as one of producing a closed tree of the transition
steps.  It is convenient to use the formal framework of {\em
  Constraint Logic Programming\/} (\emph{CLP\/})~\cite{jaffar94clp},
which we outline as follows.

The \emph{universe of discourse} is a set of terms, integers, and
arrays of integers.  A \emph{constraint} is written using a language
of functions and relations.

An \emph{atom} is of the form $p(\tilde{t})$ where $p$ is a
user-defined predicate symbol and the $\tilde{t}$ a tuple of terms.  A
\emph{rule} is of the form $p(k,\tilde{x}) ~\clpif~ p(k',\tilde{x}')
\wedge \tilde{c}$ where the atom $p(k,\tilde{x})$ is the \emph{head}
of the rule, and the atom $p(k',\tilde{x}')$ and the (conjunction of)
constraint $\tilde{c}$ (possibly relating the variables $\tilde{x}$
and $\tilde{x}'$) constitute the \emph{body} of the rule.  Here both
$k$ and $k'$ are positive numbers denoting program points or the
special constant $\dpred{error}$ to denote an error location. We may
omit either the atom or the constraint from the body. A \emph{goal}
has exactly the same format as a body of a rule. Given a goal ${\cal
  G}$: $p(k,\tilde{x}) \wedge \phi,$ we denote by $\dpred{cons}({\cal
  G})$ the constraint $\phi$ or $\dpred{true}$ when $\phi$ is empty.

Each CLP rule represents a transition in the program\footnote{For lack
  of space, we refer readers to~\cite{jaffar09intp} and its references
  for more details about the translation from transition systems to
  CLP programs.}. For example, given a program fragment with two
variables \texttt{x} and \texttt{y}, the assignment \texttt{5: x =
  y+1 6:} is represented as the rule $p(5,x,y) ~\clpif~ p(6,x',y')
\wedge y'=y \wedge x'=y+1.$ For a conditional \texttt{6: if (x>0)
 7:}, we represent the transition between $\pp{6}$ and $\pp{7}$ by
the rule $p(6,x,y) ~\clpif~ p(7,x',y') \wedge y'=y \wedge x'=x\wedge
x>0.$

A \emph{substitution\/} simultaneously replaces each variable in a
term or constraint into some expression. We specify a substitution by
the notation $[\tilde{e}/\tilde{x}],$ where $\tilde{x}$ is a sequence
$x_1,\ldots,x_n$ of variables and $\tilde{e}$ a list $e_1,\ldots,e_n$
of expressions, such that $x_i$ is replaced by $e_i$ for all $1\leq
i\leq n.$ Given a substitution $\theta,$ we write as $e\theta$ the
application of the substitution to an expression $e.$ A {\em
renaming\/} is a substitution which maps variables into variables.
A \emph{grounding\/} is a substitution which maps each variable into a
value in its domain. A \emph{ground instance} of a constraint, atom
and rule is defined in the obvious way.

Given a goal $\G\equiv p(k,\tilde{x}) \wedge \Psi(\tilde{x}),$ $\lbr\G\rbr$
is the set of the groundings $\theta$ of the primary variables
$\tilde{x}$ such that $\tilde{\exists} \Psi(\tilde{x})\theta$ holds.
A goal $\overline{\G} \equiv (k,\tilde{x}),
\overline{\Psi}(\tilde{x})$ {\it subsumes} another goal $\G \equiv
p(k',\tilde{x}') \wedge \Psi(\tilde{x}')$ if $k=k'$ and $\lbr\overline{\G}\rbr
\supseteq \lbr\G\rbr$.  Equivalently, we say that
$\overline{\G}$ is a {\it generalization} of $\G$.  We write $\Gone
\equiv \G_2$ if $\Gone$ and $\G_2$ are generalizations of each other.

We use the notion of \emph{reduction\/} to represent symbolic
strongest postcondition operation.  Let a rule $R:$ $p(k,\tilde{x})
~\clpif~ p(k',\tilde{x}') \wedge \tilde{c}$ belong to a CLP
program. Given a goal $\G:$ $p(k,\tilde{x}_i) \wedge \Psi$ with
variables disjoint from $R,$ a \emph{reduct\/} or \emph{derivation\/}
of $\G$ using $R$ (denoted $\dpred{reduct}_R(\G)$) is the goal
$p(k',\tilde{x}_{i+1}),
\Psi\wedge\tilde{c}[\tilde{x}_i/\tilde{x}][\tilde{x}_{i+1}/\tilde{x}']$.
A derivation sequence (path) is a sequence of goals $\G_0, \G_1,
\cdots $ where $\G_i, i > 0$ is a reduct of $\G_{i-1}$.

A goal $\G: p(k,\tilde{x})\wedge\tilde{c}$ is called {\em terminal\/}
if there are no applicable rules to perform reduction on it, and it is
called {\em looping\/} if it is derived from another goal with the
same $k$ (called its {\em looping parent\/}) through one or more
reduction steps. A goal is {\em infeasible\/} if its constraints are
unsatisfiable, and a derivation sequence is so called when it ends in
an infeasible goal.

\mysection{Algorithm: Minimax}
\newcommand{\calltable}{{\cal C}_{T}}
\newcommand{\successtable}{{\cal M}_{T}}
\newcommand{\depth}{{\cal D}}
\label{sec:algorithm}

As mentioned above, there is an obvious strategy for dealing with
loops by using iterative deepening on the level of loop unrolling, and
in each iteration, to generate loop invariants.  In this section, we
present an algorithm that performs unrolling in an \emph{intelligent}
manner, using information about why a particular path does not
suffice.  In this regard, there is similarity to \textsc{cegar} where,
if a candidate loop invariant is found insufficient (too weak), the
refinement process takes into account the \emph{reason} for this
insufficiency in order to arrive at the next refinement.

Our algorithm maintains knowledge about a state (goal) $\G$ $\equiv$
$p(k,\tilde{x}) \wedge c_{1}\wedge\ldots\wedge c_{n}$ by means of a
vector $v \equiv \langle \alpha^1, ... , \alpha^n \rangle$ where each
$\alpha^i$ is an \emph{annotation} of one of the following kinds:

\begin{itemize}
\item a \mymax{} annotation, indicating that the constraint $c_i$ \emph{must be kept} 
\item a \mymin{} annotation, indicating that the constraint $c_i$ \emph{must be deleted}, or 
\item a \emph{neutral} annotation.
\end{itemize}
Denote the $i$-th annotation in $v$ by $\alpha^i_v.$ Let $c$ be a
constraint, its annotation is denoted $\alpha_v(c).$
$\dpred{neut}(\tilde{c})$ denotes a vector
$\langle\dpred{neutral},\ldots,\dpred{neutral}\rangle$ of the same
length as $\tilde{c}.$ We write $\dpred{conflict}(v_1,v_2)$ if
$\exists~1\leq
i\leq\dpred{min}\{\dpred{length}(v_1),\dpred{length}(v_2)\}$ such that
($\alpha_{v_1}^i = \dpred{min}$) and ($\alpha_{v_2}^i=\dpred{max}$).

A pair $(\G,v)$ where the state
$\G \equiv p(k,\tilde{x})\wedge\tilde{c}$ is called
an \emph{annotated state}. The meaning of an annotated state $\sigma = (\G,v)$
is obtained in two ways.
A \mymax{} interpretation $\sigma_{\dspred{max}}$ is the state
obtained by deleting all but the \mymax{}-annotated constraints in
$\tilde{c}$.
Dually, a \mymin{} interpretation $\sigma_{\dspred{min}}$ is the state
obtained by including all but the \mymin{}-annotated constraints in
$\tilde{c}$.
\noindent
For example, given an annotated state $\sigma$: $p(5, x_1, x_2,
x_3) \wedge x_1 = 1\wedge x_2 = 2\wedge x_3 = 3,$ $\langle \mymin,$
$neutral,$ $\mymax \rangle,$ $\sigma_{\dspred{max}}$ and
$\sigma_{\dspred{min}}$ are, respectively the two states $p(5, x_1,
x_2, x_3) \wedge$ $x_2 = 2$ $\wedge$ $x_3 = 3$ and $p(5, x_1, x_2,
x_3) \wedge x_3 = 3.$ Note 
$\dpred{cons}(\sigma_{\dspred{min}})$ is weaker than
$\dpred{cons}(\sigma_{\dspred{max}}).$

The use of vectors is an efficient way for computing interpolants.
$\dpred{max}$-annotated constraints of an annotated state
$\sigma=(\G,v)$ must be kept to preserve some infeasible paths in the
derivation tree emanating from $\sigma$. Given an infeasible annotated
state $\sigma'$ derived from $\sigma$, we minimally
$\dpred{max}$-annotate the constraints in $\sigma'$ (some of which are
constraints of $\sigma$ since they share the vector) such that the
infeasibility is maintained. In this way we immediately obtain an
abstraction at $\sigma$ (that is, $\sigma_{\dspred{max}}$) by the
$\dpred{max}$ annotations generated at $\sigma'$ without performing
{\em weakest precondition\/} propagation or some approximation of
it. $\sigma_{\dspred{max}}$ subsumes $\G$ yet it entails the
infeasibility of $\sigma'$ and therefore is an interpolant. The final
abstraction at $\sigma$ is a conjunction of the interpolants returned
by the children, and this is easily obtained by the conjunction of all
$\dpred{max}$-annotated constraints at $\sigma$ after the subtree is
traversed.

\begin{wrapfigure}{l}{1.6in}
\input{algoutline.pstex_t}
\caption{Min-Max}
\label{fig:outline}
\vspace*{-0.5cm}
\end{wrapfigure}

The algorithm operates on annotated states.  Its depth-first traversal
is outlined in Fig.~\ref{fig:outline}.  When encountering a loop
(point $L$ in Fig.~\ref{fig:outline}) a loop invariant is produced by
weakening the constraints at $L$ by minimally $\dpred{min}$-annotating
its state. This weakening is then applied in the forward execution of
the points beyond the $L.$ This abstraction, however, is not the final
abstraction that is used to subsume other states since it still can be
weakened further as some constraints may not contribute to the
infeasibility or subsumption of descendant states. The final abstract
state (in $L$ or elsewhere) is computed by propagating $\dpred{max}$
annotations backward in post-order manner. $\dpred{Max}$ annotations
are produced by interpolation at points where infeasibility and
subsumption are found
(Lines~\ref{falseinterpolation},~\ref{conflictinterpolation}
and~\ref{subsumptioninterpolation} in Fig.~\ref{fig:minimax}).

Before detailing the algorithm of Fig.~\ref{fig:minimax} we first
explain its main components.

\vspace{0.25cm}
\noindent \textbf{Interpolation.}  If $\sigma$ is
$(p(k,\tilde{x}) \wedge \tilde{c}, v)$ and
$\dpred{cons}(\sigma_{\dspred{min}}) \wedge \varphi$ is unsatisfiable,
\textsf{interpolate}($\sigma$, $\varphi$) returns an annotation $v'$
which has the same length as $v$ (and $\tilde{c}$), satisfying the
following:
\begin{enumerate}
\item $\forall c \in \tilde{c} :
  \alpha_{v}(c)\in\{\dpred{min},\dpred{max}\} \Rightarrow
  \alpha_{v'}(c) = \alpha_{v}(c),$
\item $\forall c \in \tilde{c} : \alpha_{v}(c)=\dpred{neutral}
  \Rightarrow \alpha_{v'}(c)\in\{\dpred{neutral},\dpred{max}\},$ and
\item $\sigma' = (p(k,\tilde{x})\wedge\tilde{c},v') \Rightarrow
  (\dpred{cons}(\sigma'_{\dspred{max}}) \wedge \varphi
  ~\mbox{unsatisfiable})$
\end{enumerate}
$v'$ is computed by adding the fewest \mymax{} annotations to neutral
annotations in $v$, thus representing a computation of an {\em
  interpolant:\/} $\sigma'_{\dspred{max}}$ maintains the
unsatisfiability (consequence) of $\sigma_{\dspred{min}}$ yet it has
less constraints (more general).  For example, consider the annotated
state $\sigma:$ $(p(k, x_1, x_2) \wedge \tilde{c},$ $\langle \mymin,$
$\mymax,$ $neutral,$ $neutral \rangle)$ with $\tilde{c}$ be
$x_1>3\wedge$ $x_1 = y_1 + 1\wedge$ $y_1 = 2\wedge$ $x_2 = 0,$ and a
constraint $\varphi: x_1<0.$ where Here, $\sigma_{\dspred{min}}$ is
unsatisfiable.  Then \textsf{interpolate}$(\sigma,\varphi)$ produces the
vector $v':$ $\langle \mymin,$ $\mymax,$ $\mymax,$ $neutral\rangle.$
That is, the third constraint's annotation is changed from
\textit{neutral\/} to \textit{max} such that $\sigma'_{\dspred{max}}
\wedge \varphi$ maintains the unsatisfiability.

\vspace{0.25cm}

\noindent \textbf{Subsumption and Loop Invariants.} An essential
feature of our algorithm, existing also in \textsc{ar} methods, is the
ability of blocking the forward search traversal of an annotated state
$\sigma$ if there exists another state $\sigma'$ already processed
such that the state of $\sigma$ entails the state associated with
$\sigma'$. During the symbolic traversal there are two kinds of
subsumptions.

\emph{Parent-Child:} assume that $\sigma'$ is a looping ancestor of
$\sigma$. Here $\sigma'$ would be of the form
$(p(k,\tilde{x}')\wedge\tilde{c}_1,v_1)$ and $\sigma$ of the form
$(p(k,\tilde{x})\wedge\tilde{c}_1\wedge\tilde{c}_2,v),$ where $v = v_1
\cdot v_2$ with $v_2$ of the same length as $\tilde{c}_2.$
Since we would like to unroll as few times as possible, the algorithm
forces (if possible) parent-child subsumption by computing the
strongest path-based loop invariant. Therefore, $v_1$ can be replaced
with a vector $\overline{v_1}$ of the same length where some
\textit{neutral\/} annotations (those that are not individually
invariant) in $v_1$ are transformed to \textit{min\/} annotations in
$\overline{v_1}$ such that $\sigma'_{\dspred{min}}$ subsumes
$\sigma_{\dspred{min}}$.  The function
$\mbox{\textsf{invariant}}(\sigma,\sigma')$ returns the vector
$\overline{v_1} \cdot v_2$ if subsumption holds. Otherwise, the
parent-child subsumption is not possible and the algorithm returns
$\bot$. This our mechanism to lazily unroll loops.

\emph{Sibling-Sibling:} assume now the state $\sigma'$ has been
already processed and stored in a \emph{memo table},
$\successtable$. The condition here is that the current state
associated to $\sigma$ entails the interpolant associate to
$\sigma'$. That is, $\sigma'_{\dspred{max}}$ subsumes
$\sigma_{\dspred{min}}$. This test is done by the function
$\mbox{\textsf{subsumed}}(\successtable,\sigma)$ in the algorithm.  If
the test holds, this function also returns a \emph{subsuming state}
$\sigma^{\dspred{sub}}$. Otherwise, $\bot$.

For $\sigma^{\dspred{sub}}$ we need to distinguish two subcases. If
$\sigma'$ is out of the scope of a loop, then $\sigma^{\dspred{sub}} =
\sigma'$. Otherwise, as in the case of parent-child subsumption, we
may need to convert some \textit{neutral\/} annotations into
\textit{min\/} annotations to communicate ancestors the conditions
under the subsumption took place. In particular, those neutral
annotations which if had been \textit{max\/} annotations then
subsumption would not have held.

\vspace{0.25cm}
\noindent \textbf{Merging Vectors.} 
We use two operations for merging both \mymin{}
and \mymax{} annotations. Given two vectors $v_1$ and $v_2$:

\vspace{0.15cm}
\noindent \textsf{mergemin}($v_{1},v_{2}$): if the condition $\forall
1\leq i\leq\dpred{min}\{\dpred{length}(v_1),\dpred{length}(v_2)\} :
\alpha^i_{v_1} = \dpred{min} \Rightarrow \alpha^i_{v_2}
\in\{\dpred{neutral},\dpred{min}\}$ holds then it returns a vector $v$
satisfying $\forall 1\leq i\leq\dpred{length}(v_1):$ $(\alpha^i_{v_1}
= \dpred{min} \Rightarrow \alpha^i_v = \dpred{min}) \wedge$
$(\alpha^i_{v_1} \neq \dpred{min} \Rightarrow \alpha^i_v =
\alpha^i_{v_2}).$ Otherwise, the function returns $\bot.$

\vspace{0.15cm}
\noindent \textsf{mergemax}($v_1,v_2$): returns always a vector $v$
satisfying $\forall 1\leq i\leq\dpred{max}$ $\{\dpred{length}(v_1),$
$\dpred{length}(v_2)\}:$ $((\alpha^i_{v_1} = \dpred{max} \vee
i>\dpred{length}(v_2)) \Rightarrow$ $\alpha^i_v = \alpha^i_{v_1})
\wedge$ $((\alpha^i_{v_1} \neq \dpred{max} \vee i>\dpred{length}(v_1))
\Rightarrow$ $\alpha^i_v = \alpha^i_{v_2}).$

\newcounter{proglineno}
\setcounter{proglineno}{0}
\newcommand{\putno}{\refstepcounter{proglineno}\arabic{proglineno}:}

\begin{figure*}[t]
\hspace*{-2mm}\begin{tabular}{rl}
&  \textsf{Minimax}($\depth,$ $\sigma,$ $\calltable$) \textbf{returns} (\textsc{ok},$a$,$b$) or
(\textsc{conflict},$a$,$b$) with vector $a$ and integer $b$\\
&  \hspace{2mm} \textbf{let} $\sigma$ be $(\G,v)$ and $\G$ be $p(k, \tilde{x})\wedge\tilde{c}$\\
&  \hspace{2mm}  \textbf{switch}($\sigma$)  \\
\putno &  \hspace{5mm} \textbf{case} $\dpred{cons}(\sigma_{\dspred{min}})$ unsatisfiable: \\
\putno \label{falseinterpolation} &  \hspace{10mm} \textbf{return} (\textsc{ok},\textsf{interpolate}($\sigma$, $\dpred{true}$),0) \\
\putno &  \hspace{5mm} \textbf{case} $k = \dpred{error}$: \\
\putno \label{errorstart} &  \hspace{10mm} \textbf{if} ($\tilde{c}$ is satisfiable) \textbf{abort}\\
\putno \label{conflictinterpolation} &  \hspace{10mm} $v' ~\mbox{:=}~ \mbox{\textsf{interpolate}}((\G,\dpred{neut}(\tilde{c})),\dpred{true})$\\
\putno \label{conflictdepth} &  \hspace{10mm} $d ~\mbox{:=}~ \dpred{min}\{l|(l,(\G',v'')) \in \calltable ~\mbox{and}~ \dpred{conflict}(v'',v')\}$\\
\putno \label{errorend} &  \hspace{10mm} \textbf{return} (\textsc{conflict},\textsf{mergemax}($v'$, $v$),$d$) \\  
\putno \label{terminal} &  \hspace{5mm}  \textbf{case} $\G$ is terminal: \textbf{return} (\textsc{ok},$v$,0) \\
\putno \label{subsumingstart} &  \hspace{5mm} \textbf{case} There is $\sigma^{\dspred{sub}} = \mbox{\textsf{subsumed}}(\successtable,\sigma)$ and $\sigma^{\dspred{sub}} \neq \bot:$\\
\putno \label{subsumptioninterpolation} &  \hspace{10mm} \textbf{return} (\textsc{ok},$\mbox{\textsf{interpolate}}(\sigma,\neg\dpred{cons}(\sigma^{\dspred{sub}}_{\dspred{max}}))$,0)\\
\putno \label{loopingstart} &  \hspace{5mm} \textbf{case} $S = \{\sigma'|(l,\sigma')\in\calltable ~\mbox{and}~ \sigma' ~\mbox{looping parent of}~ \sigma\}\neq\emptyset $:\\
\putno &  \hspace{10mm} \textbf{foreach} $\sigma' \in S$\\
\putno \label{invariantdiscovery} &  \hspace{15mm} \textbf{if} ($v' = \mbox{\textsf{invariant}}(\sigma,\sigma')$ and $v'\neq\bot$) \textbf{return} (\textsc{ok},$v'$,0) \\
\putno \label{loopingend} &  \hspace{10mm} \textbf{goto} \textbf{default}\\
&  \hspace{5mm}  \textbf{default}   : \\
\putno \label{defaultstart} & \hspace{10mm} $v'$ ~\mbox{:=}~ $v$, $\sigma$ ~\mbox{:=}~ $({\cal G},v')$ \\

\putno \label{eachchild} & \hspace{10mm} \textbf{foreach} ${\cal G}'$ \textbf{in} $\dpred{red}^+_{\dspred{min}}(\sigma)$ \ldots $\dpred{red}^-_{\dspred{min}}(\sigma)$\\

\putno & \hspace{15mm} \textbf{let} $\dpred{cons}({\cal G}')$ be $\tilde{c} \wedge \tilde{c}'$\\
\putno \label{recursivecall1} & \hspace{15mm} ($\dpred{Status}$,$v''$,$d$) ~\mbox{:=}~ \textsf{Minimax}($\depth+1,$ $({\cal G}', ~v' \cdot \dpred{neut}(\tilde{c}'))$, $\calltable \cup \{(\depth,\sigma)\}$) \\
\putno \label{incrementalintp} & \hspace{15mm} $v'''$ = $\overline{\dpred{wp}}((\G',v''),\tilde{c}')$\\
\putno & \hspace{15mm} \textbf{if} ($\dpred{Status} = \mbox{\textsc{conflict}}$)  \\
\putno & \hspace{20mm} \textbf{if} ($d = \depth$)\\
\putno \label{tableclear} & \hspace{25mm} $\successtable ~\mbox{:=}~ \successtable \backslash \{\sigma'|\sigma'=({\cal G}'',\_\ignore{v^{{\i}v}}) ~\mbox{and}~ {\cal G}'' ~\mbox{derived from}~ {\cal G}\}$\\
\putno \label{recursivecall2} & \hspace{25mm} \textbf{return} \textsf{Minimax}($\depth,$ $(\G, v''')$, $\calltable$) \\
\putno \label{forwardconflict} & \hspace{20mm} \textbf{else return} (\textsc{conflict},$v'''$,$d$) \\
\putno \label{minabstract} & \hspace{15mm} $v'$ ~\mbox{:=}~ \textsf{mergemax}($v'''$,\textsf{mergemin}($v''', v'$)) \\
\putno \label{memoing} & \hspace{10mm} $\successtable ~\mbox{:=}~ \successtable \cup \{(\G,v')\}$\\
\putno \label{defaultend} & \hspace{10mm} \textbf{return} (\textsc{ok},$v'$,0)
\end{tabular}
\caption{The Minimax Algorithm}
\label{fig:minimax}
\vspace*{-0.5cm}
\end{figure*}

The {\sf Minimax} routine takes as inputs the depth $\depth$ of the
symbolic tree, a current annotated state $\sigma$, and the table
$\calltable$ to record the ancestor states that can potentially become
the looping parent of the current state.  There is a global table,
$\successtable$, to store the interpolants already computed.  The
execution starts with some $\depth=0$, $\sigma_{\dspred{init}}$ which
is neutral, and an empty $\calltable$.  The memo table,
$\successtable$, is also initially empty.

\indent Line~\ref{falseinterpolation} handles the case when the state
is infeasible. Here, \mymax{} annotations are created using the
procedure \textsf{interpolate} to indicate constraints that are needed
in order to preserve unsatisfiability of the constraints.

\indent Lines~\ref{errorstart}--\ref{errorend} handle the case when
\textit{error\/} program point is visited with feasible
$\sigma_{\dspred{min}}.$ In case $\sigma$ is itself feasible
($\tilde{c}$ is satisfiable), we have found a real error, and the
algorithm aborts (Line~\ref{errorstart}).  In case $\sigma$ is
infeasible, we have found a spurious state, which is visited due to
the weakening caused by $\dpred{min}$ annotations. At
Line~\ref{conflictinterpolation} we compute $\dpred{max}$ annotations
such that the infeasibility is preserved. At Line~\ref{conflictdepth}
we compute the shallowest depth value such that the conflict occurs,
from which we are to restart. We then add the computed $\dpred{max}$
annotations to the input vector and returns the resulting vector
together with a \textsc{conflict} status and the computed depth
(Line~\ref{errorend}).

\indent Line~\ref{terminal} is selected if the end of the path is
reached.  Here it is not necessary to add either $\dpred{min}$ or
$\dpred{max}$ annotations as looping points or infeasible/subsumed
states can no longer be reached, and we therefore return \textsc{ok}
and the input annotation itself.

\indent Lines~\ref{subsumingstart}--\ref{subsumptioninterpolation}
handle the case if the current state is subsumed by another state
already memoed in $\successtable$. Recall the notion of subsuming
state $\sigma^{\dspred{sub}}$ explained previously. Here we return
\textsc{ok} with a vector with more $\dpred{max}$ annotations than the
input vector $v$ needed to ensure the entailment (unsatisfiability of
the negation of the constraints) of the abstraction of the subsuming
state ($\sigma^{\dspred{sub}}_{\dspred{max}}$).

\indent Lines~\ref{loopingstart}--\ref{loopingend} handle the case
when the current state is looping. Here we attempt to compute a
path-based loop invariant using $\dpred{min}$ annotations that are
produced by calling \textsf{invariant} subprocedure
(Line~\ref{invariantdiscovery}) in order to force parent-child
subsumption.  If \textsf{invariant} fails to produce the abstraction,
we continue to the default case (Line~\ref{loopingend}).

\indent The default case at Lines~\ref{defaultstart}--\ref{defaultend}
performs one symbolic execution step.  We first formalize functions
used in Line~\ref{eachchild}.  Let $\sigma$ be $(\G,v)$, we denote by
$\dpred{red}^+_{\dspred{min}}(\sigma)$ the set $\{\G'|\exists R :
\dpred{cons}(\dpred{reduct}_R(\sigma_{\dspred{min}})) ~\mbox{is
  satisfiable}~ \wedge \G'=\dpred{reduct}_R(\G)\}.$ Similarly, we
denote by $\dpred{red}^-_{\dspred{min}}(\sigma)$ the set
$\{\G'|\exists R :
\dpred{cons}(\dpred{reduct}_R(\sigma_{\dspred{min}})) ~\mbox{is
  unsatisfiable}~ \wedge \G'=\dpred{reduct}_R(\G)\}.$ In essence,
$\dpred{red}^+_{\dspred{min}}(\sigma)$
($\dpred{red}^-_{\dspred{min}}(\sigma)$) is the set of reducts of $\G$
such that, using the same rules, the reduction of
$\sigma_{\dspred{min}}$ is feasible (infeasible). In this way, the
loop in Line~\ref{eachchild} makes us prioritize transitions that are
feasible, possibly due to $\dpred{min}$ abstraction.  This is
important to not generate $\dpred{max}$ annotations that restricts
abstraction too early resulting in failure to discover loop invariants
later (inability to convert $\dpred{max}$ to $\dpred{min}$ annotations
at Line~\ref{invariantdiscovery}). Then, the loop iterates in sequence
over the reducts,
performing recursive calls to \textsf{Minimax}
(Line~\ref{recursivecall1}).  The result, for each one, is a triple
($\dpred{Status},v'',d$).  $v''$ here contains $\dpred{max}$
annotations that specify how the current state need to be abstracted.
At Line~\ref{incrementalintp} we compute the abstraction for the
current state based on the annotation returned using the function
$\overline{\dpred{wp}}$, which denotes an approximation of the weakest
precondition. In our framework, this can be trivially done, without
calling the theorem prover, by cutting off the last $|\tilde{c}'|$
elements of $v''$. (Here $\tilde{c}'$ are the constraints added by the
reduction.)

If $\dpred{Status}$ is \textsc{conflict} with depth $d$ (produced at
Line~\ref{errorend}), we know that somewhere during the recursive
call, a conflict occurred. If the depth $d$ is equal to the current
depth $\depth$, then this is the topmost point where the conflict is
originated. In addition, we know that $v'''$ (the vector after calling
$\overline{\dpred{wp}}$) is the same as $v$, but with some
$\dpred{max}$ annotations replacing $\dpred{min}$ annotations. In
essence, we ``lock'' such annotations. More importantly, this may
result in failure to create loop invariant at
Line~\ref{invariantdiscovery} later. At Line~\ref{tableclear} we need
to clean the memo table for those states derived from the current
input state.  We then perform another recursive call
(Line~\ref{recursivecall2}) as a replacement for the current call
(without making any transition step), and using $v'''$ to propagate
the locked annotations. If the current depth is not equal to the
conflict depth, we simply propagate the conflict to the parent
(Line~\ref{forwardconflict}).

\indent Finally, Line~\ref{minabstract} combines the vectors returned
by each descendant, and after all vectors are merged,
Line~\ref{memoing} stores in the memo table the interpolant for the
current goal.

We conclude this section by mentioning that the central step of
deleting constraints, the effect of a \mymin{} annotation, can in fact
be relaxed to some other mechanism that abstracts the state at hand.
Instead of deleting a constraint, one could \emph{transform} a
constraint.  For example, one could apply a process of ``slackening''
to equations $x = y$ to obtain an inequality, either $x \leq y$ or $x
\geq y$.  This kind of abstraction is in fact employed in the
\textsc{blast} system which we benchmark against, but at this time, we
do not use for our own experimental results.  Even more generally, we
could replace not one but a collection of constraints by another
collection which is entailed by the original collection.

\mysection{Experimental Evaluation}
\label{sec:results}

We implemented our prototype \textsc{tracer} modelling the C heap
using the theory of arrays with alias analysis to partition and
inlining functions. We ran \textsc{tracer} on several programs
instrumented with safety properties and compare with
\textsc{blast}~\cite{beyer07blast}\footnote{We tried with
  \textsc{armc} \ignore{available at~\cite{ARMC-web}} but we were only
  successful to run on \texttt{tcas} and \texttt{statemate} but
  timeout expired in both cases after 30m and 1h, respectively.}.  We
downloaded all programs from~\cite{BLAST-web} already instrumented
with safety conditions, and together with a script which runs those
programs with the most favorable system options.

The results are summarized in Table~\ref{tab:result-1}. We present two
sets of numbers: for \textsc{blast} the number of discovered
predicates (\textsf{P}) the total time in seconds (\textsf{T}), and
for \textsc{tracer}, our prototype tool, the number of nodes of the
exploration tree (\textsf{S}) and also the total time in seconds
(\textsf{T}). Although the number of discovered predicates and nodes
of the exploration tree are not comparable they are shown to provide
an idea about the hardness of the proof.

\begin{wrapfigure}{l}{6.5cm}
\vspace{-1cm}
\begin{small}
\begin{center}
  \begin{tabular}{|l|r||c|c||c|c|} \hline & &
    \multicolumn{2}{c||}{\textsf{\textsc{blast} (\textsc{ar})}} &
    \multicolumn{2}{c|}{\textsf{\textsc{tracer} (\textsc{al})}} \\ \cline{3-6}
    \textsf{Program} & \textsf{LOC} & \textsf{P} &
    \textsf{T} & \textsf{S} & \textsf{T} \\
    \hline\hline
\texttt{qpmouse} & 400 & 4  &  0.42 &  974  &  0.42  \\ 
\hline
\texttt{tlan} &  8069 &   14  &  17.10 &  4382 & 5.78  \\ 
\hline 
\hline
\texttt{cdaudio}&  8921  &  * &  *    &  6258  &  10.53  \\ 
\hline
\texttt{diskperf}&  6984 & 92 &  82.3 &  3326  &  8.21  \\ 
\hline
\texttt{floppy}&  8570   &  * &  *    &  3124  &  6.47  \\ 
\hline
\texttt{kbfiltr}& 5931 &  45   &  44.03    &  1392   & 2    \\ 
 \hline 
\texttt{serial}& 10380 
                                & *   &  *         &  59597      &  328.6     \\ \hline 
\hline
\texttt{tcas-1a-safe}  & 394 & 23 & 3.6    & 6029  & 6.97 \\ 
\texttt{tcas-1b-safe}  &     & 56 & 78.35  & 6050  & 6.77 \\ 
\texttt{tcas-2a-safe}  &     & 22 & 3.25   & 6029  & 6.74 \\ 
\texttt{tcas-3b-safe}  &     & 39 & 15.68  & 6017  & 6.63 \\ 
\texttt{tcas-5a-safe}  &     & 31 & 10.29  & 6029  & 6.36 \\ 
\texttt{tcas-2b-unsafe}&     & 40 & 17.46  & 91    & 0.01 \\
\texttt{tcas-3a-unsafe}&     & 25 & 18.96  & 243   & 0.16 \\ 
\texttt{tcas-4a-unsafe}&     & 45 & 14.44  & 243   & 0.15 \\ 
\texttt{tcas-4b-unsafe}&     & 36 & 6.44   &  91   & 0.01 \\ 
\texttt{tcas-5b-unsafe}&     & 54 & 40.31  &  91   & 0.02 \\ \hline 
\hline
\end{tabular}
\caption{\textsc{blast} Benchmarks on Intel 2.33Ghz 3.2GB}
\label{tab:result-1}
\end{center}
\end{small}
\vspace{-1cm}
\end{wrapfigure}

In summary, \textsc{tracer} is competitive with \textsc{blast} in most
of the benchmark examples, sometimes much faster. However, there are
two programs where \textsc{blast} is faster
(\texttt{tcas-1a\ignore{-safe}}, and
\texttt{tcas-2a\ignore{-safe}}). We believe the main reason is that
\textsc{tracer} does perform some extra work due to unnecessary
infeasible paths. Nevertheless, the numbers show that the differences
are not significant.

Note that programs such as \texttt{cdaudio}, \texttt{floppy}, and
\texttt{serial} are annotated with the symbol '*' in the
\textsc{blast} column which means that \textsc{blast} raised an
exception and aborted. Therefore, we were not able to verify those
programs using \textsc{blast}. Although we could not contact 
\textsc{blast} authors we are aware that \texttt{cdaudio} and
\texttt{floppy} have been proved safe in~\cite{henzinger04proof} after
21m59s and 11m17s discovering 196 and 156 predicates, respectively on
Pentium 2.4Ghz 512Mb.

Special mention deserves the cases where the programs were proved
unsafe. In these cases, \textsc{tracer} found a real counterexample
much faster than \textsc{blast}. The reason is that \textsc{tracer}
blocks infeasible paths and then finds very quick the real
error. \textsc{blast} will spent some time performing refinements and
traversing space which are irrelevant to the real error
path. Nevertheless, this is an example where we believe that
Synergy-like tools using test cases would perform as ours since
\textsc{dart} could also find the real error path faster.

\mysection{Concluding Remarks}
\label{sec:conclusions}

We extended Abstraction Learning, an interpolation-based symbolic
execution method, to automatically handle unbounded loops.  The
algorithm is an intelligent unrolling process by classifying into
\mymin{} and \mymax{} constraints. The \mymin{} constraints are those
which must be abstracted in order to achieve subsumption and loop
invariance, while the \mymax{} constraints are those which must not be
abstracted so as to detect infeasible paths and also to preserve
safety.  The idea is to have as few of these two kinds of constraints
as possible.  We discussed the relative merits of ours and
\textsc{ar}-based methods using academic examples.
We also evaluated our prototype, \textsc{tracer}, against
\textsc{blast}, the most advanced system available to us, using real
programs.  The results show competitive performance, with some
examples showing significant improvement. In all cases, the results
show that eagerly detecting infeasible paths can be efficient.


{\small
\bibliographystyle{plain}

 }

\end{document}